\documentclass[12pt,a4paper]{article}
\usepackage{amsfonts,amssymb,amsmath}
\usepackage{epsfig}
\usepackage{graphicx}

\title{A nonstationary form of the range refraction parabolic equation and its application
 as an artificial boundary  condition for the wave equation in a waveguide}

\author{Pavel Petrov\thanks{ E-mail: petrov@poi.dvo.ru},\,  Mikhail Trofimov\thanks{ E-mail: trofimov@poi.dvo.ru}\\ 
Il'ichev' Pacific Oceanological Institute,\\ Vladivostok, Russia}

\begin{document}

\maketitle

\begin{abstract}
The time-dependent form of Tappert's range refraction
parabolic equation is derived and proposed as an artificial boundary
condition for the wave equation in a waveguide. The numerical
comparison with Higdon's absorbing boundary conditions shows
sufficiently good quality of the new boundary condition at low
computational cost.
\end{abstract}

\section{Introduction}
In this paper we consider wave propagation in a two-dimensional waveguide $\Omega = \{(x,y)|\, a<y<b\}$
described by the wave  equation
\begin{equation} \label{t1}
L\,u =
\frac{1}{c^2} \frac{\partial^2u}{\partial t^2}  - \frac{\partial^2u}{\partial x^2}
 - \frac{\partial^2u}{\partial y^2}  = 0\,,
\end{equation}
where the sound speed $c$ is a function of spatial variables, $c=c(x,y)$. For various purposes
one needs to obtain  one way or unidirectional wave equations, which permit wave propagation only
in  certain directions along $x$-axis. For example, for such equations we can consider $x$ as
the evolution variable and set initial boundary value problems with initial data at $x=x_0$,
which are not well posed for the wave equation itself.
\par
The main approach to this problem  consists in factorization of the
wave operator in eq.~(\ref{t1}) into  unidirectional
pseudodifferential factors
\begin{equation} \label{t11}
L  =  L^+\cdot L^- \,,\quad
L^\pm =
\frac{\partial}{\partial x} \pm D\,,\quad D=
\sqrt{\left(\frac{1}{c^2} \frac{\partial^2}{\partial t^2} - \frac{\partial^2}{\partial y^2}\right)}\,,
\end{equation}
where  the positive square root is used. Then various approximations
to the square root are used for obtaining concrete unidirectional
equations. Note that the factorization in eq.~(\ref{t11}) is not
exact when $c$ depends on $x$, so this approach requires optional
assumptions of the asymptotic character on this dependence.
In fact,
$$
L  =  L^+\cdot L^- + \left[D\,,\,\partial/\partial x\right] =
L^+\cdot L^- -D_x\,,
$$
so it is sufficient to assume that $c$ depends on $x$ slowly,
$c=c(\delta x,y)$ for some small parameter $\delta$. Then $D_x =
O(\delta)$.
 In this paper we also assume that $c$ is regular enough with respect to $y$ to perform
 the calculations below.
\par
In the approximation of the square root in eq.~(\ref{t11}) the
assumption of small angle propagation with respect to the $x$-axis
is usually used, i.~e. the operator $(1/c^2)(\partial^2/\partial
t^2)=A$ is treated as $O(1)$ and $(\partial^2/\partial y^2)=\epsilon
B$ is $O(\epsilon)$ in some small parameter $\epsilon$. In the case
of constant sound speed $c$, when the operators $A$ and $B$ commute,
the application of  usual rational approximations to the square root
function then yields local (differential) unidirectional wave
equations.
Under the assumption that $X=A+\partial^2/\partial y^2-K$ is small for some constant $K$,
the usual rational approximations with respect to $X$ can also be applied to approximate
the square root operator and its exponential (one-step propagator). This approach is well developed for the
time-harmonic case \cite{c-hong,c-hong2,c-hong3}.

\par
 With the strong dependence of $c$ on $y$ the operators $A$ and $B$ do not commute and the applicability
 of the standard rational approximations to the square root operator in eq.~(\ref{t11}) is unclear.
 In fact, difficulties arise already in consideration of the first order (linear in $B$) approximation.
 As we know now, this problem in a particular case was first solved correctly by R.~Feynman,
  who in his paper \cite{feyn} derived  the formula
\begin{equation} \label{t12}
\mathrm{e}^{A+\epsilon B}=\mathrm{e}^{A} +
\epsilon\int_0^1\mathrm{e}^{(1-s)A} B \mathrm{e}^{sA}\,ds + O(\epsilon^2)\,.
\end{equation}
\par
In the late 1970s, F.~Tappert in his paper \cite{tapp} derived the
so called range refraction parabolic equation for the time-harmonic
sound propagation in waveguides with arbitrary dependence of the
index of refraction on depth:
\begin{equation} \label{t1T}
\mathrm{i}v_x + \frac{1}{2k_0}\frac{\partial}{\partial y}\left(\frac{1}{n}\frac{\partial v}{\partial y}\right)
+k_0\left[n+\frac{1}{4k_0^2}\left(\frac{n_y^2}{n^3}-\frac{n_{yy}}{n^2}\right)\right] v =0\,,
\end{equation}
where $n$ is the index of refraction and $k_0$ is the reference wave number.
He used the expansion of the operator
square root in the form very similar to eq.~(\ref{t12}) (see
eq.~(\ref{t3}) below), but without any derivation. As Feynman's paper
was also not cited, the derivation of eq.~(\ref{t1T}) in Tappert's paper seemed to be somewhat
mysterious.
\par
In the next section we derive the nonstationary form of the Tappert's equation, using the same
expansion, but derived from some elementary results of noncommutative analysis \cite{nazai}.
We hope that this information will be useful for the reader.
\par
Note that Tappert's equation (\ref{t1T}) is of restricted use for
the computation of the wave field because it is not an {\em
amplitude} equation. The same is true for our new eq.~(\ref{n7}). As
an application of the new equation we consider its use as an
artificial  (absorbing, non-reflecting) boundary condition (see,
 e.g.,  \cite{hagstrom,hagstrom2,eng-maj,tsynkov,antoine}) at the boundaries $x=x_0$. These
artificial boundaries and the corresponding boundary conditions are
needed for restricting the computational domain to simulate waves
propagating freely in the whole waveguide. Such an use is possible
because any unidirectional equation partially annihilates waves
propagating in the opposite direction.
\par
Note that an analogous problem where the variability of coefficients
is also essential was studied by a different method in \cite{krieg,antoine2}.

\section{Derivation of the time-dependent form of Tappert's range refraction parabolic equation}
Due to the equality
$$
\int_0^1 f'(sx+(1-s)y)\,ds = \frac{f(x)-f(y)}{x-y}
$$
the first order term in eq.~(\ref{t12}), denote it $C$, can be
written in the form
$$
C= \epsilon \stackrel{2}{B}
\frac{\mathrm{e}^{\stackrel{3}{A}}-\mathrm{e}^{\stackrel{1}{A}}}{\stackrel{3}{A}-\stackrel{1}{A}}\,,
$$
where the numbers above operators (now called the Feynman numbers) define the order
in which they are to operate \cite{feyn,nazai}. ``Thus, $BA$ may be written $\stackrel{2}B\stackrel{1}A$ or
$\stackrel{1}A\stackrel{2}B$" \cite{feyn}.
The same formula holds for some general class of operator functions $f$
\begin{equation} \label{t4}
f(A+\epsilon B) = f(A) +
\epsilon\stackrel{2}{B}\frac{f(\stackrel{1}{A})-f(\stackrel{3}{A})}{\stackrel{1}{A}-\stackrel{3}{A}}+
O(\epsilon^2)\,,
\end{equation}
as was proved few years later by Daletskiy and Krein (see \cite{nazai}).
\par
Using eq.~(\ref{t4}), the operator $C$ in the expansion
\begin{equation} \label{t21}
\left(A+\epsilon B\right)^{1/2} = A^{1/2} + \epsilon C +O(\epsilon^2)
\end{equation}
can be directly computed:
\begin{equation}\label{t3}
\begin{split}
&C=\stackrel{2}{B}\frac{\stackrel{1}{A}^{1/2}-\stackrel{3}{A}^{1/2}}{\stackrel{1}{A}-\stackrel{3}{A}}
= \stackrel{2}{B}\frac{1}{\stackrel{1}{A}^{1/2}+\stackrel{3}{A}^{1/2}} \\
&=\stackrel{2}{B}\int_0^\infty \mathrm{e}^{-s \stackrel{1}{A}^{1/2}}
\mathrm{e}^{-s \stackrel{3}{A}^{1/2}}\,ds   =
\int_0^\infty \mathrm{e}^{-s A^{1/2}} B
\mathrm{e}^{-s A^{1/2}}\,ds\,.
\end{split}
\end{equation}
This formula can be verified  by squaring both sides of
eq.~(\ref{t21}), which gives the equation for $C$
$$
A^{1/2}C + C A^{1/2} = B\,.
$$
Substitution into the last formula the expression for $C$ from
eq.~(\ref{t3}) immediately gives
$$
A^{1/2}C + C A^{1/2} =
-\int_0^\infty \frac{d}{ds}\left(\mathrm e^{-s A^{1/2}}B \mathrm e^{-s A^{1/2}}\right)\,ds =
B\,.
$$
The last equality and the representation (\ref{t3}) holds because
\begin{equation} \label{n7aa}
\mathrm{e}^{-s A^{1/2}} u = \exp \left({-s \frac{1}{c}\frac{\partial}{\partial t}}\right) u
= u(t-\frac{1}{c}s,x,y)\,,
\end{equation}
and the functions under consideration vanish with their derivatives
before some time moment, say, for $t<0$.

\par
By a straightforward calculation using eq.~(\ref{n7aa}) we first obtain
\begin{equation} \label{n7a}
\begin{split}
&Cu = \int_0^\infty \frac{(c_y)^2}{c^4}\cdot s^2\cdot u_{tt}(t-2\frac{1}{c} s,x,y)\,ds \\
&+ \int_0^\infty \left(\frac{c_{yy}}{c^2} - 2\frac{(c_y)^2}{c^3}\right)\cdot s\cdot u_t(t-2\frac{1}{c} s,x,y)\,ds \\
& + \int_0^\infty u_{yy}(t-2\frac{1}{c} s,x,y)\,ds \\
&+ 2\int_0^\infty u_{yt}(t-2\frac{1}{c} s,x,y)\cdot s \frac{c_y}{c^2} \,ds\,.
\end{split}
\end{equation}
Then, using the formula
$$
u_{t}(t-2\frac{1}{c} s,x,y) = -\frac{c}{2}\frac{\partial}{\partial s}u(t-2\frac{1}{c} s,x,y)
$$
and the analogous formula for $u_{tt}$, we eliminate  the time derivatives
and $s$ in eq.~(\ref{n7a}) by integration by parts as follows
\begin{equation*}
\begin{split}
&\int_0^\infty  s^2\cdot u_{tt}(t-2\frac{1}{c} s,x,y)\,ds \\
&= c\int_0^\infty s\cdot u_t(t-2\frac{1}{c} s,x,y)\,ds \\
&= \frac{c^2}{2}\int_0^\infty  u(t-2\frac{1}{c} s,x,y)\,ds \,.
\end{split}
\end{equation*}
 At last, changing the variable $s$ by $s_{new} = 2s_{old}/c$, we find
  the final expression for the
operator $C$ and so the time-dependent form of Tappert's equation
for  right- and left-propagating waves
\begin{equation} \label{n7}
\begin{split}
u_x \pm \frac{1}{c}u_t \mp \frac{1}{4}\left(c_{yy}-\frac{c^2_y}{c}\right)\int_0^{\infty}u(t-s,x,y)ds \\
\mp \frac{1}{2}\int_0^{\infty}\left(c u_y(t-s,x,y)\right)_y\,ds =
0\,.
\end{split}
\end{equation}
\par
Differentiating eq.~(\ref{n7}) with respect to $t$ using the above technique, we obtain its differential form
\begin{equation} \label{n7d}
u_{xt} \pm \frac{1}{c}u_{tt} \mp \frac{1}{4}\left(c_{yy}-\frac{c^2_y}{c}\right)u
\mp \frac{1}{2}\left(c u_y\right)_y = 0\,,
\end{equation}
which is, of course, not completely equivalent to eq.~(\ref{n7}),
especially in computational aspects. Eq.~(\ref{n7}) is more easily
handled and, due to its nonlocality, has better stability properties
than eq.~(\ref{n7d}).
\par
It is easy to check that for the time harmonic $u=\mathrm{e}^{\mathrm{i}\omega t}v(x,y)$ eq.~(\ref{n7d})
coincides with the original range refraction parabolic equation for the field $v$ (\ref{t1T}).
Conversely, the Fourier transform with respect to time of  eq.~(\ref{t1T}),
multiplied by $\mathrm{e}^{\mathrm{i}\omega t}$, gives eq.~(\ref{n7d}) and then, after integration with respect
to $t$, eq.~(\ref{n7}).

\section{Eq.~(\ref{n7}) as an artificial boundary condition}
In this paper we will apply eq.~(\ref{n7}) for left going waves as an artificial boundary
condition on the artificial boundary $x=x_0$, the computational domain is located in $x>x_0$.
As was mentioned in the Introduction, the problem consists in simulating waves, propagated in the unbounded
waveguide, by solving some initial boundary value problem in the restricted computational domain.
\par
As far as we know, only the technique of perfectly matched absorbing layers
can be immediately applied to the problem under consideration in the case
of waveguides with variable sound speed $c$ (see \cite{hagstrom}).
Besides this technique  the most frequently used by the practitioners boundary conditions
have the form
\begin{equation} \label{giv1}
\left[\prod\limits_{j=1}^{J}\left(\frac{\partial}{\partial
t}+C_j\frac{\partial}{\partial x}\right)\right]u=0\,,
\end{equation}
where $C_j$ are some constants.
 In the paper \cite{higdon} Higdon proved (in the case of constant $c$) that  the class of these conditions
 contains all boundary conditions obtained by the Pad\'e approximations of the operators
 $L^\pm$ (the Engquist-Majda boundary conditions \cite{eng-maj}).  It is immediately seen that the condition (\ref{giv1})
perfectly annihilates  waves of the form $f(Ct-x)\phi(y)$ if the
phase speed $C$ is equal to some $C_j$. The boundary conditions
(\ref{giv1}) are referred in the literature as the Higdon conditions of
order $J$.
\par
The main problem in any practical use of the Higdon conditions consists in finding
some (quasi)optimal set  $\{C_j\}$.
For waves with given frequency spectrum it is possible to try the phase or group velocities
of the corresponding normal modes. The automatic choice, described in \cite{giv-net2}, also gives
good results for dispersive waves.
In essentially non-stationary problems for the wave equation the correct choice of $C_j$ is difficult and
the optimal set $\{C_j\}$ may depend on time. So in such ambiguous situations the use of some mean
value of the sound speed $c$ as $C_j$ is probably more preferable.
\par
As our boundary condition reduces to the second order Engquist-Majda condition
in the constant sound speed case, we expect that it will work better than the
Higdon condition of order 2 in the variable sound speed case. The performed numerical experiments
confirm this conclusion.
\par
We must make some remarks on the well-posedness of the introduced boundary condition.
In the case of constant
sound speed one can recognize in eq.~(\ref{n7d}) the second order
Engquist-Majda boundary condition \cite{eng-maj}, which is proved to
lead to a well-posed mixed initial boundary value problem for the
wave equation. The proof consists in checking the {\em uniform
Kreiss condition} (U.K.C.) and is valid as well in the variable
coefficient case (see, e.g., \cite{higdon2}). As the third term in
eq.~(\ref{n7d}) is unessential for the U.K.C., we may conclude that
the boundary condition (\ref{n7d}) leads to a well-posed mixed
initial boundary value problem for the wave equation. As
eq.~(\ref{n7}) implies eq.~(\ref{n7d}), the same is true for the
boundary condition (\ref{n7}).

\section{Finite difference discretization}

In our numerical experiments we have used
the standard second-order explicit finite difference scheme for the wave equation on the uniform grid
$\{t^i,x_j,y_l\}$
\begin{equation} \label{n5}
\begin{split}
\frac{1}{c^2_{j,l}}D^+_tD^-_tu^i_{j,l}=D^+_xD^-_xu^i_{j,l}+D^+_yD^-_yu^i_{j,l},
\end{split}
\end{equation}
where $u^i_{j,l}=u(t^i,x_j,y_l)(t^i=i\tau,x_j=jh-h/2,y_l=lh,i=0,1\dots ,j=1,2\ldots,l=1,2\ldots)$,
$D^+_xu^i_{j,l}=(u^i_{j+1,l}-u^i_{j,l})/h$,
$D^-_xu^i_{j,l}=(u^i_{j,l}-u^i_{j-1,l})/h$, and $D^+_t, D^-_t, D^+_y,
D^-_y$ are defined analogously.

\par
 We assume that $u$ and all its derivatives vanish for all
 $t<0$. Under this assumption the boundary condition of the
Tappert type (\ref{n7}) on the left boundary at $x=0$ are
discretized as:
\begin{equation} \label{n8}
\begin{split}
D^+_x(u^{i}_{1,j}+u^{i+1}_{1,j})/2-D^+_t(u^{i}_{1,j}+u^{i}_{2,j})/(2c_{3/2,j})+\\
+\frac{\tau}{8}\left(c_{yy}-\frac{c^2_y}{c}\right)_{3/2,j}\cdot
\sum\limits^i_{m=2}(u^{m}_{1,j}+u^{m+1}_{2,j})+\\
+\frac{\tau}{8}(c_y)_{3/2,j}\sum\limits^i_{m=2}D^+_yD^-_y(u^{m}_{1,j}+u^{m+1}_{2,j})+\\
+\frac{\tau
c_{3/2,j}}{4}\sum\limits^i_{m=2}D^+_yD^-_y(u^{m}_{1,j}+u^{m+1}_{2,j})\,,
\end{split}
\end{equation}
where $(\cdot)_{3/2,j}$ is the value of $(\cdot)$ at $x=0,y=jl$.
It should be noted that in practice the sound speed $c$ is often
given by some interpolation formula, so there is no need to
calculate its derivatives by the difference methods. The Higdon boundary conditions were discretized as described in
\cite{giv-net}. The implementation of the boundary condition (\ref{n7}) in the finite element framework
 can also  be easily obtained.

\section{Numerical experiments}

\begin{figure}[t]\center
    \includegraphics[width=0.7\textwidth]{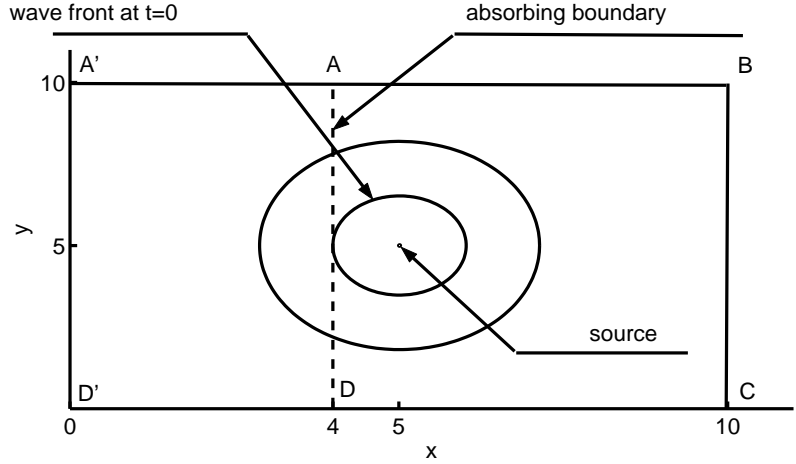}\\
  \parbox[t]{0.95\textwidth}{\caption{Computational domain for the point source. }}\label{fig4}
\end{figure}
\suppressfloats[b]
The nondimensional computational domain for these experiments is shown in fig.~1. It is a rectangle $\{x,y|\, 0<x<l_x=10\,; \,0<y<L_y=10\}$. At each boundary of the extended domain A'BCD' the hard wall (zero
Neumann) boundary conditions are applied but  waves do not reach
these boundaries during calculations. At the boundary AD of the
truncated domain ABCD the absorbing boundary conditions are imposed.
Thus solving the initial boundary value problem for the wave
equation first on the truncated domain and then on the extended one
we can estimate the quality of solutions using error measures
\begin{equation} \label{n9}
E(t_i)=
\frac{\sum\limits_{j}\sum\limits_{k}|u^i_{j,k}-u_{E}(t_i,x_j,y_k)|}
{\sum\limits_{j}\sum\limits_{k}|u_{E}(t_i,x_j,y_k)|}
\end{equation}
and
\begin{equation} \label{n10}
e(t_i)=\max\limits_{j}\max\limits_{k}|u^i_{j,k}-u_{E}(t_i,x_j,y_k)|\,,
\end{equation}

where $u$ is the solution of initial boundary value problem on the
truncated domain, and $u_E$ is the solution from the extended
domain.

\begin{figure}[t]
    \includegraphics[width=0.49\textwidth]{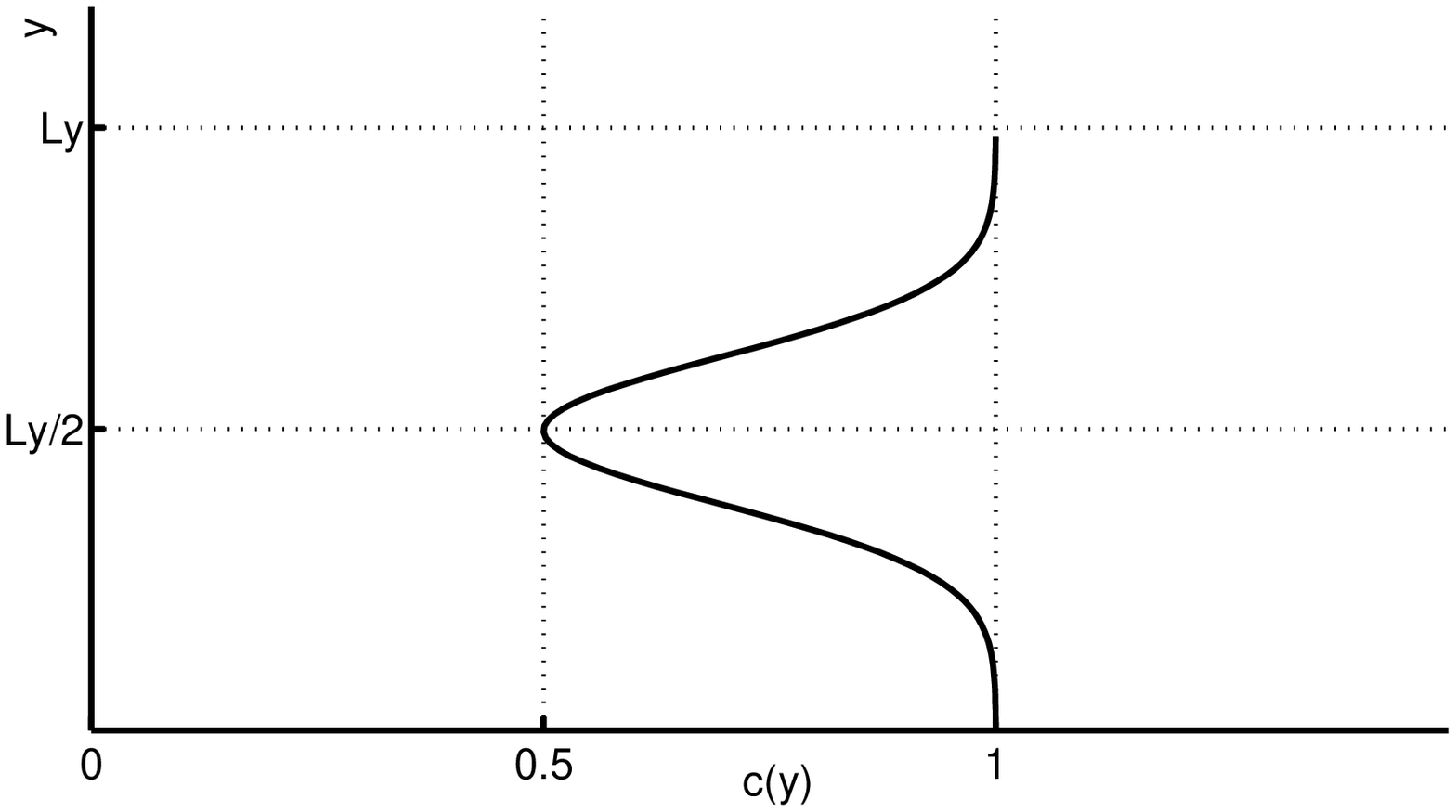} \hfill
    \includegraphics[width=0.49\textwidth]{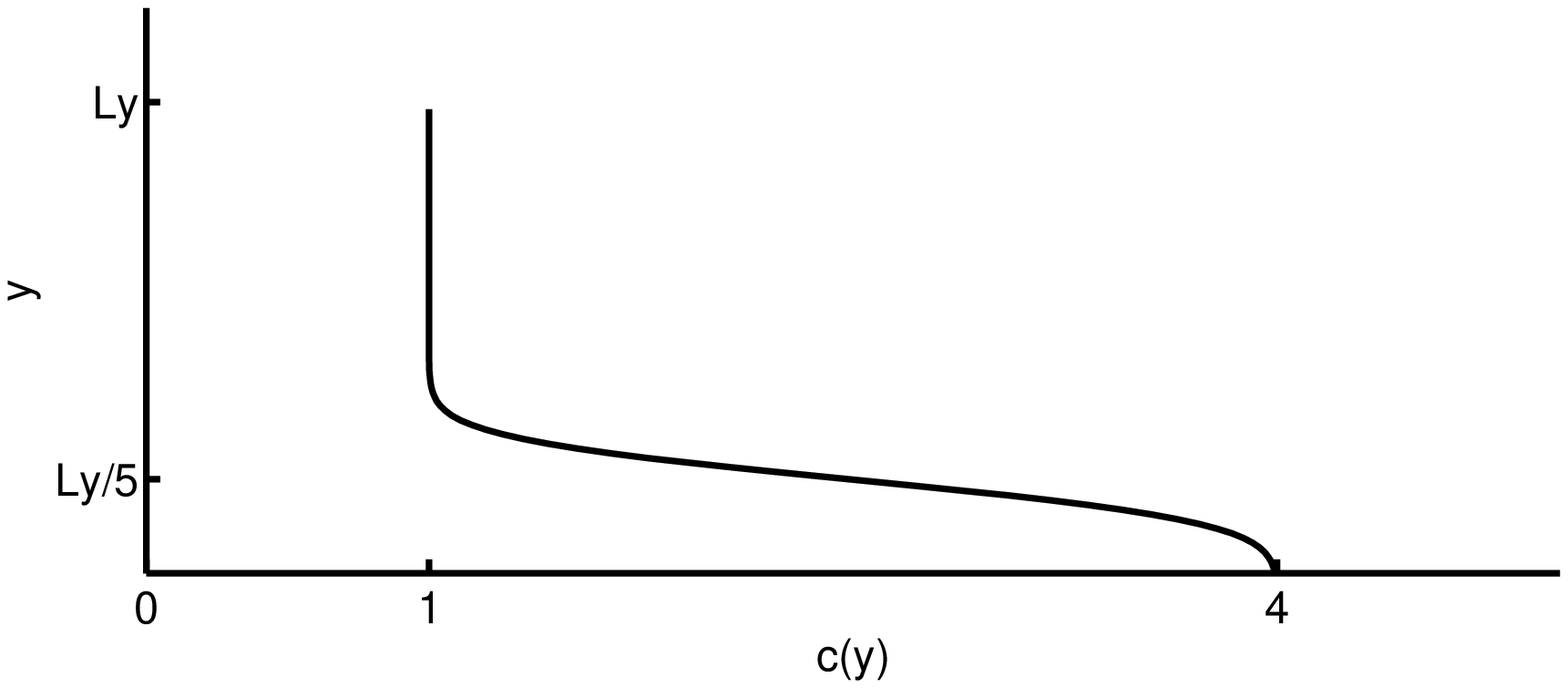}\\
  \parbox[t]{0.45\textwidth}{\caption{c(y) given by (\ref{n12}). }}\hfill
  \parbox[t]{0.45\textwidth}{\caption{c(y) given by (\ref{n14}). }}\label{ss}
\end{figure}
\suppressfloats[b]

As the initial conditions was taken the wave field in the
homogeneous media produced by the point source of short duration
located at the center of the extended domain.

\par
Numerical solutions were obtained on the grid with the space step
$\sigma=0.1$ and the sufficiently small time step $\tau$ which
guarantees stability of the finite-difference scheme.
\begin{figure}[t]
    \includegraphics[width=0.49\textwidth]{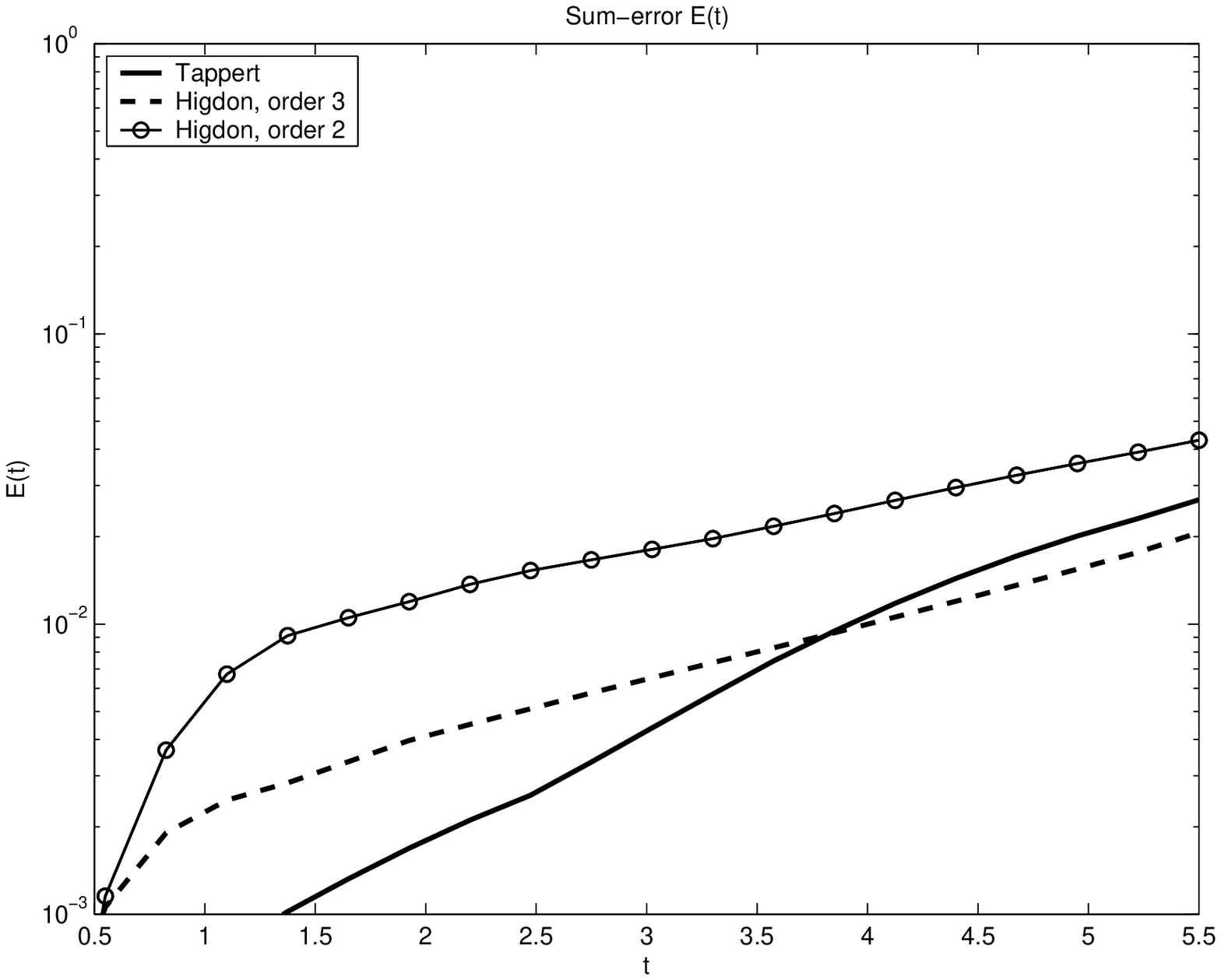} \hfill
    \includegraphics[width=0.49\textwidth]{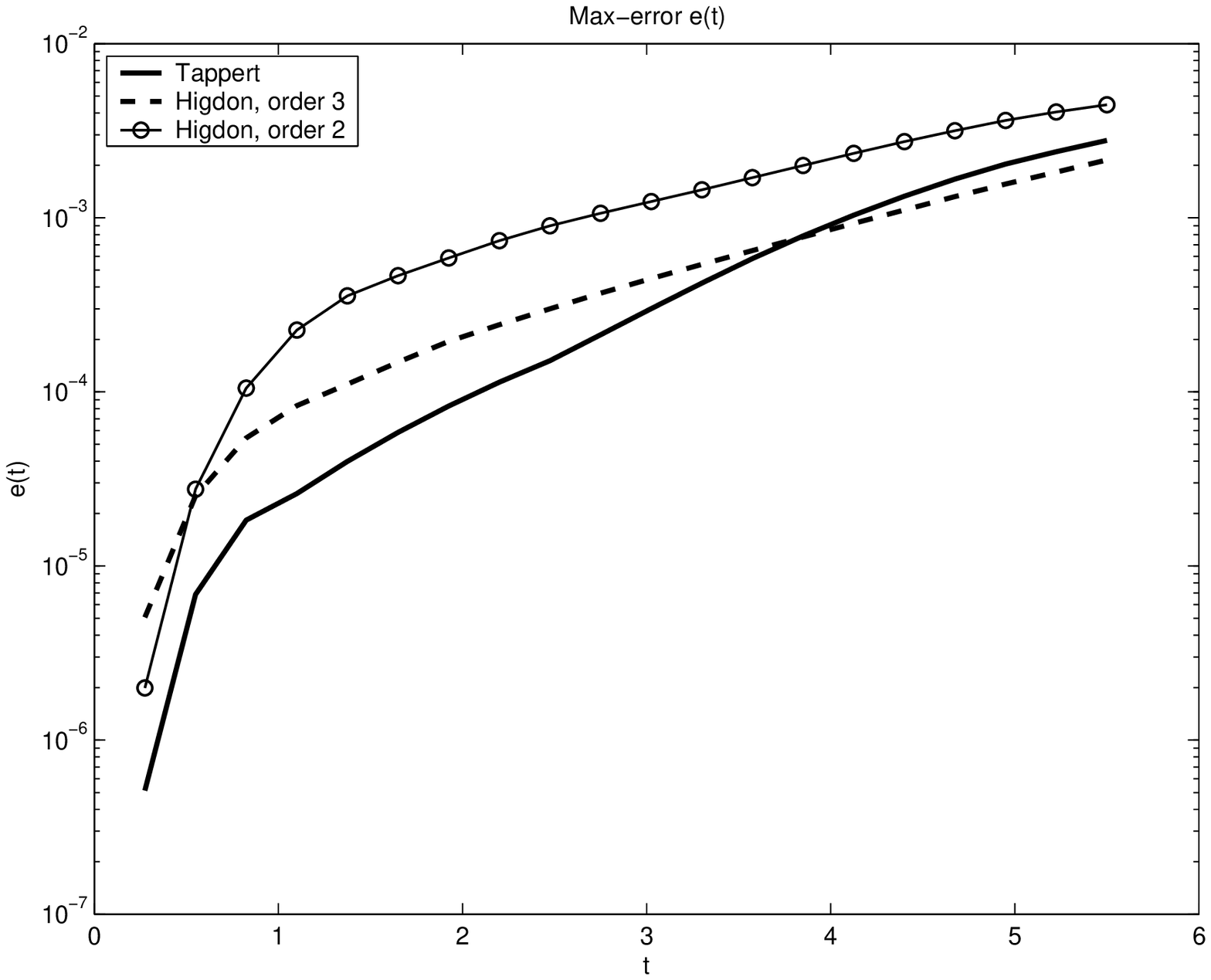}\\
  \parbox[t]{0.45\textwidth}{\caption{Errors $E(t)$ for the first example }}\hfill
  \parbox[t]{0.45\textwidth}{\caption{Errors $e(t)$ for the first example }}\label{gauss1}
\end{figure}
\suppressfloats[b]
\par
In our experiments we compare the new boundary condition, in the
sequel called the Tappert condition, with the Higdon boundary
conditions of order $J=2$ and $J=3$, with the $\{C_j\}$ given by
$$
C_j=\frac{\int_0^{L_y}c(y)\,dy}{L_y}\,.
$$
\par
For the first example we use $c(y)$ with a minimum at the depth
$L_y/2$ given by (fig.~2)
\begin{equation} \label{n12}
c(x)=1-0.5e^{(y-L_y/2)^2/3}\,,
\end{equation}

which gives $C_j\approx 0.8565$. The point source duration was taken
to be $1.4$. The results of calculations are presented in
fig.~4-5. 
In the second example the variation of
$c(y)$ is larger (fig.~3), where $c(y)$ is
defined through
\begin{figure}
    \includegraphics[width=0.49\textwidth]{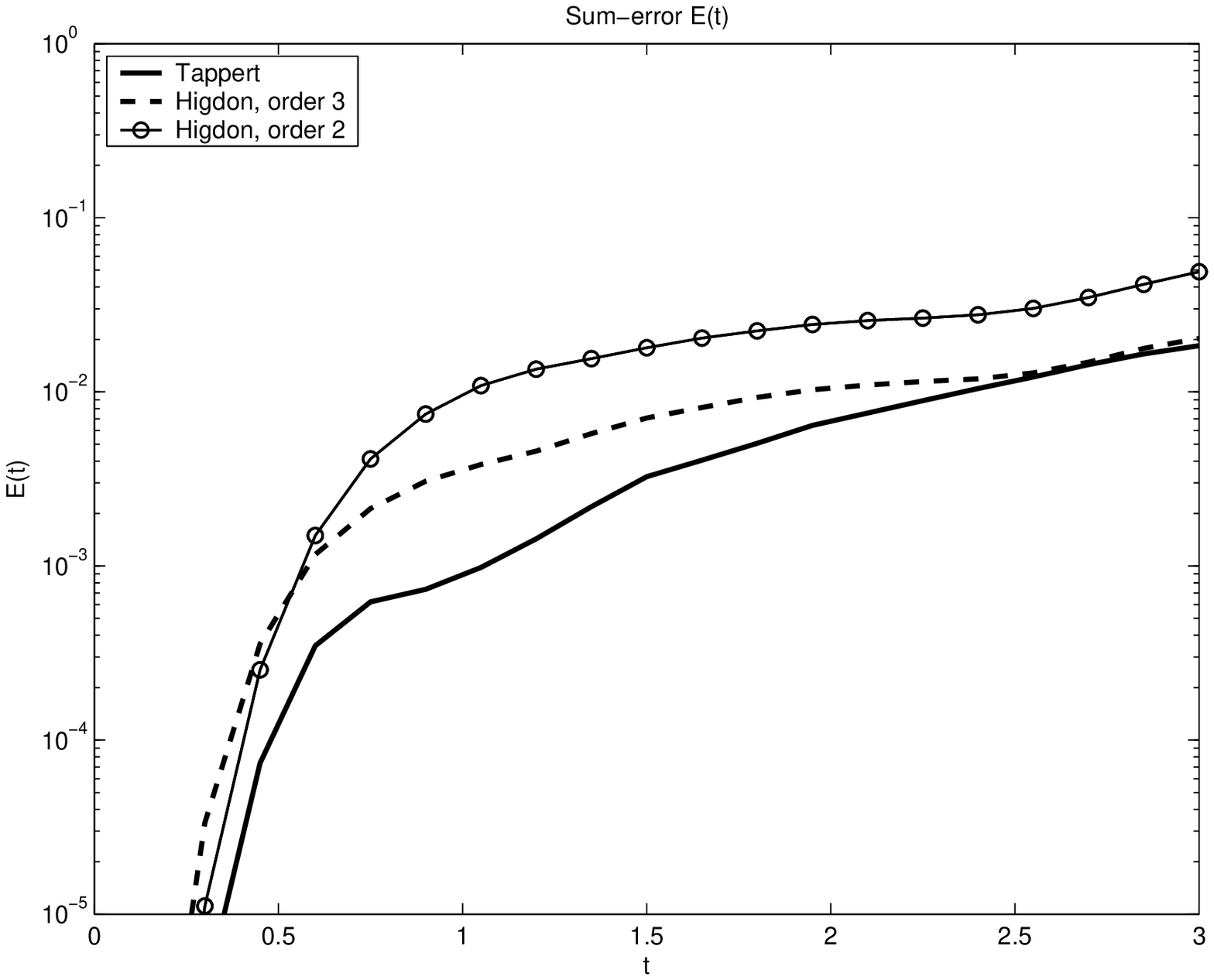} \hfill
    \includegraphics[width=0.49\textwidth]{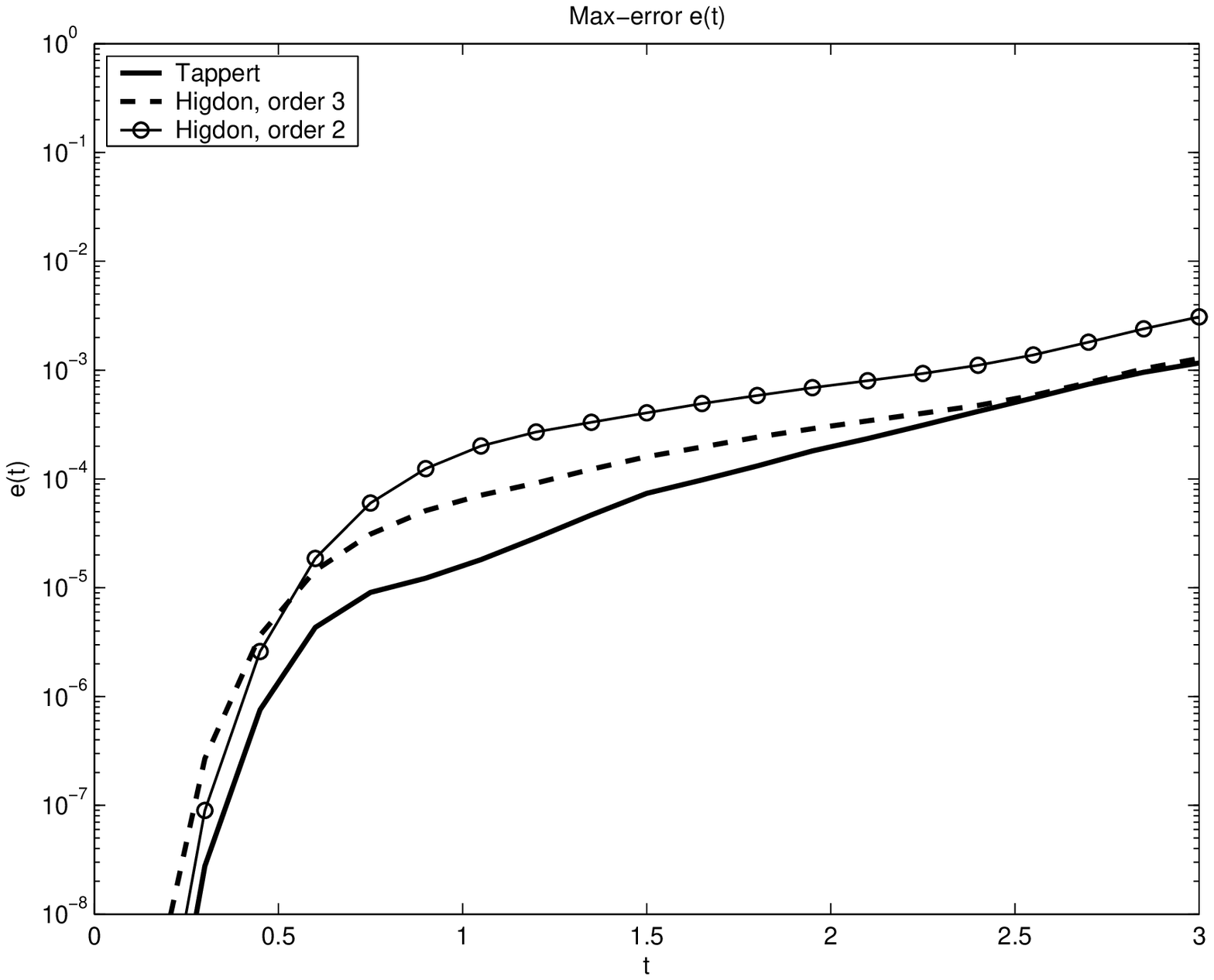}\\
  \parbox[t]{0.45\textwidth}{\caption{Errors $E(t)$ for the second example }}\hfill
  \parbox[t]{0.45\textwidth}{\caption{Errors $e(t)$ for the second example }}\label{erf1}
\end{figure}
\begin{equation} \label{n14}
c(y)=4-\frac{3}{\sqrt{\pi}}\int\limits_{-\infty}^{y}e^{(s-L_y/5)^2}\,ds\,.
\end{equation}
In this example $C_j\approx 1.6521$ and the point source duration
 was taken to be 1.
The corresponding results are presented in fig.~6-7. We see that the larger the variation of the sound speed is, the
better the Tappert condition works.
\par
 As the Tappert condition (\ref{n7}) contains no
derivatives of the sound speed  with respect to the range variable, it
can be applied in the range-dependent environment without any
modifications. If the third example the following
range-dependent sound speed $c(x,y)$ was taken:

\begin{equation} \label{n122}
c(x,y)=1-0.5e^{(x-0.7L_x)^2/3}\,.
\end{equation}

In this example the source was positioned at $x=3L_x/4,\,y=L_y/2$.
The results are presented in fig.~8-9. This and various other experiments show that in
the range-dependent case the growth of errors under the Tappert condition
(\ref{n7}) is essentially  the same  as in the range-independent
examples.

The discrete form of the Tappert boundary condition (\ref{n8}) shows that
the integrals in the  eq.~(\ref{n7}) are simply accumulated at each time step,
so the computational cost of the new boundary condition only slightly differs
from those of the first order Higdon condition. The computational cost of the second order
and the third order Higdon conditions is about 15\% and 40\% greater. These estimates
are very approximate because the calculations were conducted in the MATLAB framework.
\par
At last we should note that the behavior of the new boundary condition over long times
is essentially as good (or bad) as those of the Higdon conditions.
\begin{figure}[t]
    \includegraphics[width=0.49\textwidth]{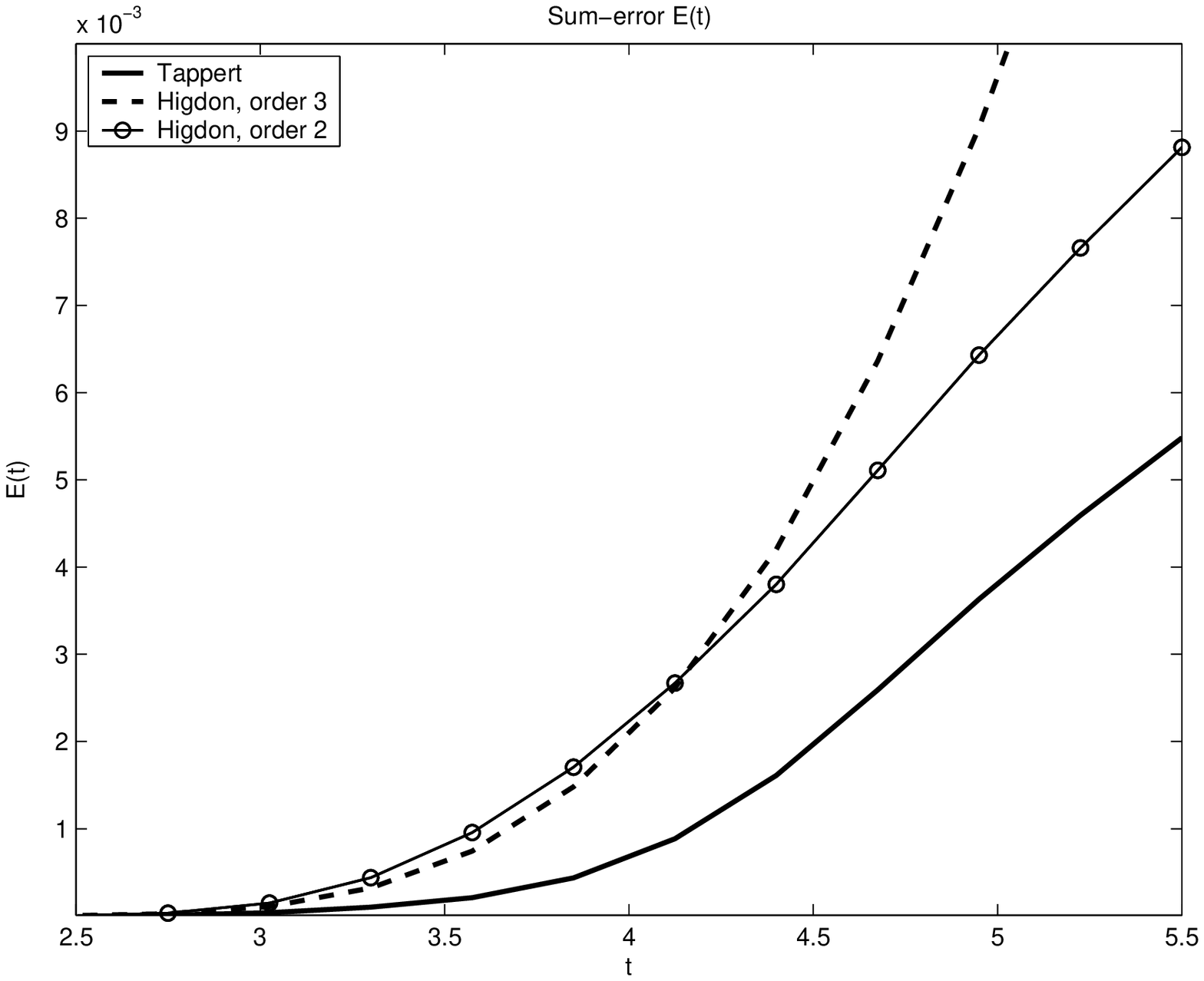} \hfill
    \includegraphics[width=0.49\textwidth]{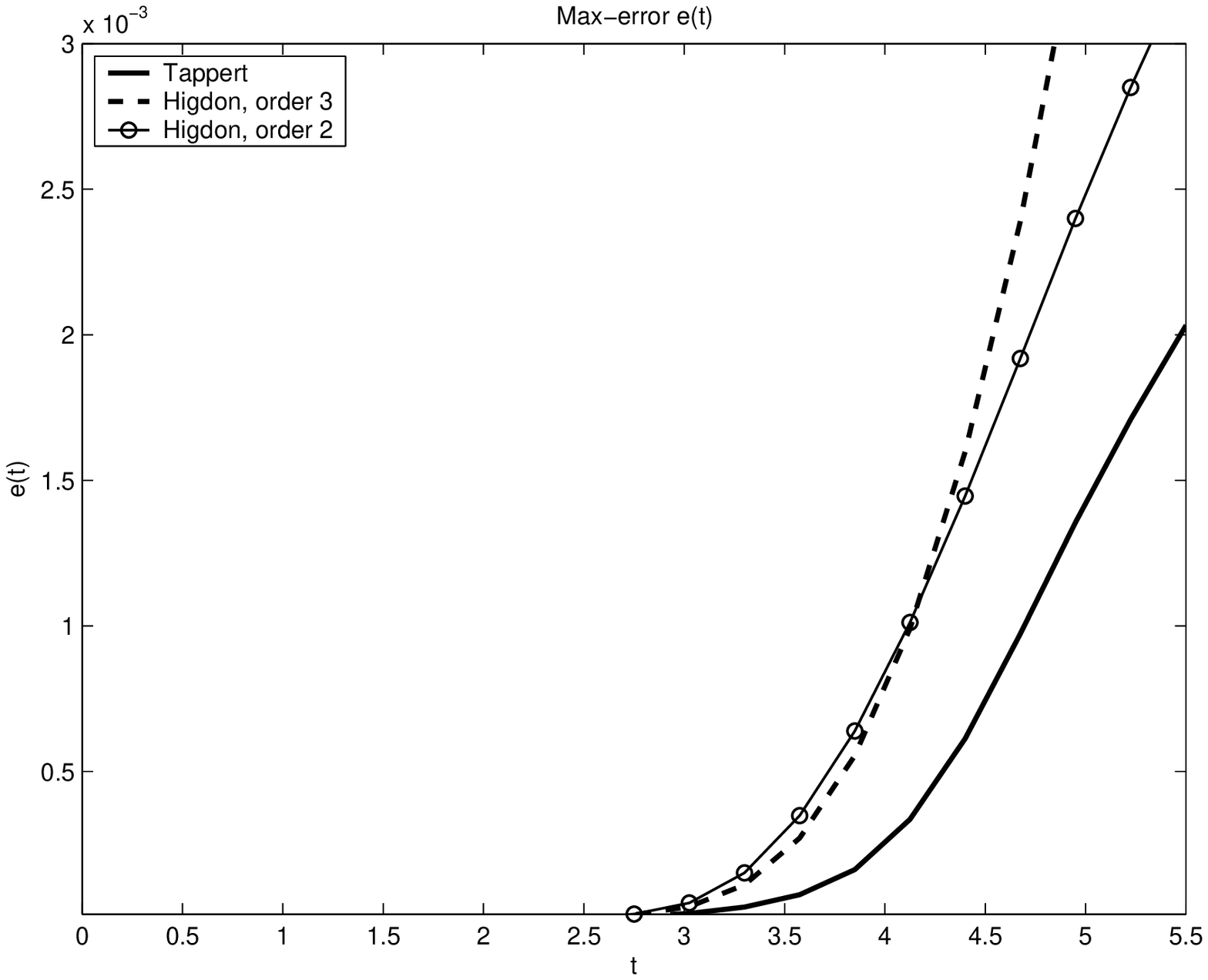}\\
  \parbox[t]{0.45\textwidth}{\caption{Errors $E(t)$ for the third example }}\hfill
  \parbox[t]{0.45\textwidth}{\caption{Errors $e(t)$ for the third example }}\label{ran}
\end{figure}

\section{Conclusion}
In this paper the time-dependent form of Tappert's range refraction
parabolic equation (\ref{n7}) is derived. This equation
describes unidirectional, small angle (with respect to the $x$-axis)
wave propagation and can be used for solving various problems of nonstationary
wave propagation which are, in particular,
ill-posed for the wave equation (\ref{t1}). Despite its nonlocality in time,
it is easily calculated.
\par
Here the new equation was applied as an  artificial boundary condition for the
wave equation in a stratified waveguide.
 Numerical experiments, performed for  comparison of the new boundary condition
with Higdon's absorbing boundary conditions, showed its sufficiently good quality  at low computational cost.

\par
Eq.~(\ref{n7}), regarded as an artificial boundary
condition, can be considered as the first member of a new sequence
of boundary conditions which are suitable for stratified waveguides.
 This sequence can be obtained by the calculation of the
following terms in expansion of the square root operator into the
Newton series (noncommutative analog of the Tailor series, see
\cite{nazai}). To achieve good stability properties the
regularization of obtained in such a way partial sums will be
necessary. This can be done  by the application of some form of
noncommutative rational approximation (e.~g., Pad\'e appoximation
\cite[Part~2,Sec.~1.5]{bak-gr}) and
 will lead also to the {\em wide angle} extensions of eq.~(\ref{n7}).

\par
  The problems for further investigations consist in the calculation of the following
members of this sequence, verification their well-posedness and automatization of their numerical
implementation. We intend  to consider some of these problems in our future studies.  The results of our
future studies will be compared with the results of the recent works \cite{hag-giv, hag-giv2} on the Higdon boundary
conditions, where the problem of variable coefficients was also considered.

 This work was supported by the Presidium of
the Russian Academy of Sciences (Programme No.~14, Part 2).

\end{document}